# Surprises from the search for quark-gluon plasma?*
## When was quark-gluon plasma seen?

Richard M. Weiner

Laboratoire de Physique Théorique, Univ. Paris-Sud, Orsay and

Physics Department, University of Marburg

The historical context of the recent results from high energy heavy ion reactions devoted to the search of quark-gluon plasma (QGP) is reviewed, with emphasis on the surprises encountered. The evidence for QGP from heavy ion reactions is compared with that available from particle reactions.

## 1. Evidence for QGP in the Pre-SPS Era

*Particle Physics*

*1.1 The mystery of the large number of degrees of freedom*

*1.2 The equation of state*

    1.2.1 *Energy dependence of the multiplicity*

    1.2.2 *Single inclusive distributions*

        *Rapidity and transverse momentum distributions*

        *Large transverse momenta*

    1.2.3 *Multiplicity distributions*

*1.3 Global equilibrium: the universality of the hadronization process*

    1.3.1 *Inclusive distributions*

    1.3.2 *Particle ratios*

---

*) Invited Review

<div align="center">* * *</div>

In February 2000 spokespersons from the experiments on CERN's Heavy Ion programme presented "compelling evidence for the existence of a new state of matter in which quarks, instead of being bound up into more complex particles such as protons and neutrons, are liberated to roam freely…"[1].

In April 2005 the four main experimental groups of the Relativistic heavy ion accelerator of Brookhaven RHIC presented their peer reviewed White papers [2] in which they summarized the results achieved in the first four years of running the RHIC accelerator. Although important new and surprising facts had been found, the RHIC reports were much more prudent in their conclusions with respect to the main goal of heavy ion reactions, the search for quark-gluon plasma (QGP): they agreed with the CERN statement only in the sense that a state of dense matter had been observed in heavy ion reactions, which was difficult to explain in terms of hadronic degrees of freedom. What has caused this unexpected step backwards?

The purpose of this paper is to review the historical context of some of the recent experimental observations at RHIC and to analyze the reasons and significance of this surprising difference of perspectives.

It will emerge that the history of the search for QGP is full of other, for some even more startling surprises. Among other things it will be reminded that the first evidence for QCD matter in the laboratory has been reported from *particle* reactions, and that already in 1979, only four years after its existence was predicted.

Independent evidence for QGP was subsequently found, again in particle reactions, in 1986 and possible indications for this new state of matter may have been found also in low energy heavy ion reactions at Bevalac and Dubna.

Moreover, evidence for quark matter in particle reactions has probably been seen already commencing with the 1950-s, although not recognized as such but in the 1970-s. These facts will be the subject of Section 1 - the Pre-SPS era. Section 2 devoted to the SPS-RHIC era, contains surprises as well. One consists in the fact that one of the main justifications of the CERN assessment is missing in the accompanying paper. Another is that in the interpretation of certain RHIC data important theoretical and experimental results obtained at SPS energies have apparently been overlooked. This presumably lead to the so called HBT puzzle "found" at RHIC energies; it will be suggested that this apparent puzzle is probably an artifact of unjustified theoretical assumptions.

The second part of section 2 discusses the historical background of the main surprise found at RHIC, which consists in the important discovery that QGP at present energies is still strongly interacting. It will be shown that this result confirms experimentally predictions based on much older phenomenological observations in particle physics, which had lead to the conjecture that hadronic matter has superfluid properties and which later on were used to describe the effect of confinement on a quark-gluon system at finite temperature. Furthermore it will be argued that it is mainly the strongly interacting property of the quark matter seen at RHIC, which explains the difference in perspectives between the two laboratories.

**Evidence for QGP in the Pre-SPS Era.**

*Particle Physics*

The evidence for QGP in particle physics is based on the success of the Landau hydrodynamical model in explaining various experimental data with an equation of state reflecting the transition from a quark-gluon plasma to a hadronic system. Besides that, the very fact that hydrodynamics is applicable presupposes a large number of degrees of freedom in the early stages of the evolution of the system. This again can be explained by assuming that in the initial phase the system consists of quarks and gluons. Last but not least, hadron production processes in hadron and lepton induced reactions are characterized by a universal production process, which suggests a common intermediate phase.

*The mystery of the large number of degrees of freedom*

The phenomenological success of the Landau hydrodynamical model [3] of multiparticle production has constituted for decades a challenge for high energy physics, because hydrodynamics is a classical theory and it assumes local equilibrium. Both these assumptions imply a large number of degrees of freedom, and in 1953, when Landau formulated his model, the only known particles were pions and nucleons, and the mean multiplicities in cosmic rays were, for present standards, quite low. Therefore the Landau model (and other statistical models of strong interactions) were considered up to the mid seventies as exotic approaches, outside mainstream physics. The bigger the merit of a few physicists like Carruthers, Feinberg, Minh,

Shuryak and others who kept the interest in the Landau model alive. (For a review of the Landau model cf. Ref. [4] and the Round Table Discussion "The Landau Hydrodynamical Model in Ref. [5]) In the West a paper by Carruthers with the provocative title "Heretical Models of Particle Production" [6] has been particularly instrumental in this situation, because it pointed out that most of the multiparticle production data obtained at the ISR accelerator could be explained by this model at least as well as by other, more fashionable approaches, like multiperipheral or parton models. Another important contribution of this paper was the distinction between "prematter" and hadronic matter [a].

The issues of local equilibrium and number of degrees of freedom in particle physics, which had constituted for decades the main conceptual difficulties of the hydrodynamical model [b], [7], got in the view of some researchers, including the present author, a simple and convincing solution with the discovery of quantum chromodynamics (QCD) and the possible existence of a new state of matter - quark gluon plasma (QGP). At present energies in a nucleon-nucleon collision, instead of 4 degrees of freedom (proton, neutron, spin up and down) one can expect 12 $N_f$ degrees, where $N_f$ (the number of flavors of quarks) ranges from 2 for u and d quarks to 3 for u, d and s quarks. 12 comes from 2 (spin) x 3 (color) x 2 (quarks and antiquarks); in addition the gluons have 2 spin degrees of freedom and 8 color degrees, so that we have some 40 to 52 internal degrees of freedom. Another circumstance, which has facilitated the acceptance of hydrodynamical methods, which belong to classical physics, was the finding that QCD admitted classical solutions [c]. Last but not least it was realized that the only chance of proving in the laboratory the existence of QGP as a *state of matter* was to use hydrodynamics in the interpretation of data. This follows from the trivial, but sometimes forgotten, fact that a state of matter is defined by an equation of state (EOS) and there is no other way to get information about the EOS than by using hydrodynamics. While the search and study of QGP in heavy ion reactions has become, with the advent of dedicated accelerators and detectors, mainstream physics [d], [8], the situation is not so clear in particle physics. That does not justify the neglect or the overlooking of possible evidence for QGP from particle physics, the more so that this evidence preceded that obtained with heavy ions.
The remainder of this section will discuss this evidence, which is based on the one hand on the form of the equation of state that follows from the comparison of the predictions of the Landau model with data, and on the other on the amazing success of the assumption of global equilibrium - an assumption much stronger than local equilibrium - in explaining the universality of the hadronization process in particle reactions. Both these facts obviously confirm the existence of a large number of degrees of freedom in strong interaction particle physics.

---

[a] Carruthers introduces the notion of "prematter," as the "medium of highly compressed and energetic hadronic matter" to be distinguished from "matter, the stuff of which S-matrix theory is composed (the asymptotic region)".

[b] Objections to the use of statistical or thermodynamical methods based on distinctions between the concepts of "phase space dominance" and "true" thermodynamical behavior are not pertinent anyway. Entropy is by definition a statistical concept and from it temperature is derived. Therefore, except for fluctuations, there is no difference between the concepts mentioned above (cf. e.g. Ref. [7], for a recent discussion of this topic).

[c] Actually the applicability of classical physics depends on the action and in QCD the vacuum plays an important role in the action, transforming systems, which look microscopic, into macroscopic ones.

**1.2.** *The equation of state*

*1.2.1 Energy dependence of the multiplicity*

According to the Landau-Pomeranchuk Ansatz, during the hydrodynamical expansion the number of particles N is not fixed. Therefore the chemical potential

$$\mu = 0$$

and

$$\varepsilon + p = TS/V.$$

Here $\varepsilon$ is the energy density, p the pressure, T the temperature, S the total entropy and V the volume. To proceed further one has to know the equation of state. Landau postulated an ideal gas EOS corresponding to a massless pion gas

---

[d] To reach this stage the "public opinion" in nuclear and particle physics had to undergo a process of learning and understanding, which has not finished yet (cf. footnote [b]). The importance and the progress of this process is reflected also in the series of special meetings dedicated to local equilibrium in strong interaction physics LESIP[8] and, of course, in the Quark Matter meetings.

$$p = \varepsilon/3 \qquad (1.1)$$

which leads to

$$S/V \sim T^3 \sim \varepsilon^{3/4}.$$

The number density is given by

$$n = N/V \sim \int d^3p / [(\exp(p^2 + m^2)/T) - 1]$$

wherefrom, in the high temperature limit, when masses are negligible, one gets

$$n \sim T^3 \sim S/V.$$

Integrating this relation we obtain

$$N \sim S \sim \varepsilon^{3/4} V.$$

Assuming that the hydrodynamical expansion of the system is adiabatic the entropy is conserved and the total number of particles N becomes defined, at a fixed freeze-out temperature $T_f$, through the last relation. To get the dependence of the multiplicity N on the energy one still has to estimate the (energy dependent) volume of the system. While the initial transverse dimension R of the system is, according to Landau, energy independent, the longitudinal size is Lorentz contracted, so that

$$V \approx 2\pi R^2 / \gamma$$

with $\gamma = E_{cm}/m_p$, where $m_p$ is the proton mass. This leads to the well known relation

$$N \sim (E_{cm})^{1/2} \sim (E_{lab})^{1/4},$$

which, up to energies of $E_{cm} = 540$ GeV, was found to be in good agreement with data. (At 540 GeV a deviation from this relationship has apparently been observed, which could be explained by the energy dependence of the inelasticity, i.e. the ratio between the energy spent on particle production and the total energy.[9]

### 1.2.2 *Single inclusive distributions*

*Rapidity and transverse momentum distributions*

In a first approximation we have a factorization [e] between the transverse momentum and rapidity distributions of the form

$$Ed^3\sigma/d^3p \approx f(p_T)g(y,s)$$

where $E_{cm} = \sqrt{s}$.

The rapidity distribution has the form

$$g \approx \exp(-y^2/2L)/(2\pi L)^{1/2}$$

with $L = \ln(\sqrt{s}/m_p)$. We see thus that in the Landau model there is no plateau in rapidity: The Landau model predicts a Gaussian rapidity distribution g with a width L, which slowly increases with energy. This prediction is also in agreement with data, up to the highest (Tevatron) energies available so far, and one might expect that this will also be true at the LHC.

The transverse momentum distribution is given by the simple formula

$$f(p_T) \sim \exp(-B\, p_T)$$

where B is independent of the energy. This is the characteristic exponential $p_T$ distribution corresponding to thermal equilibrium at an effective freeze-out temperature $T = 1/B$, taken by Landau equal to the pion mass. It is known that this formula describes well the data up to transverse momenta of ~ 1 GeV, (for a recent review of this issue cf. e.g. Ref. [10]). Moreover, there is no convincing and equally simple, alternative explanation for this experimental observation [f],[11], which in some sense is direct evidence for thermal equilibrium in strong interactions, albeit the issue of flow is not yet settled (cf. also footnote [dd]).

At this point we have to take stock of the situation. The results obtained so far show that the Landau model with an EOS of the form (1.1) successfully explains the most salient experimental observations and this suggests that Landau's educated guess about the EOS was correct. Actually it was the

---

[e] This factorization, which resembles but is not identical with boost invariance, is a consequence of the initial conditions postulated by Landau: in a head-on collision the pressure along the longitudinal direction is much larger than in the transverse direction (and therefore, initially, the transverse momentum does not increase with energy). In heavy ion collisions at AGS, SPS, and presumably also at RHIC energies, the initial conditions appear to be different, but there is no boost invariance either. At SPS an interplay between longitudinal and transverse expansion takes place, right from the beginning. This is seen, among other things, in Bose-Einstein correlations.
[f] Cf. however recent attempts in this direction in [11].

*simplest* Ansatz one could make if one considered the high energy limit where all masses are negligible and it is a fortunate accident that the EOS of QGP coincides with that of an ideal gas. The correctness of Landau's conjecture is even more amazing if one takes into account that the EOS of *hadronic* matter, the only matter known at that time, is not that represented by Eq. (1.1): If one writes the EOS under the more general form

$$p = u^2 \varepsilon, \qquad (1.2)$$

where u can be considered as an effective velocity of sound, assumed for the moment to be constant, then as shown in Ref.[12], for a gas of hadron resonances, one could rather expect

$$u^2 \approx 1/6 - 1/7. \qquad (1.3)$$

And even more to the point, we know now that the velocity of sound changes with $\varepsilon$, because of the phase transition from QGP at high temperatures to hadronic matter at freeze-out. In other words the correct velocity of sound u, which enters Eq. (1.2), is not a constant but rather a function of the energy density [g] $\varepsilon$ and Eq. (1.2) should rather read

$$p = u^2(\varepsilon) \varepsilon \qquad (1.2')$$

In the pre-QGP era this state of affairs was, of course, not known and because of that, and also because of simplicity reasons, in the applications of the hydrodynamical model always a constant velocity of sound was assumed, in most cases the canonical value $1/\sqrt{3}$.

The explanation why this constant value for u worked so well in p–p reactions was given in Ref. [13], where the (one-dimensional) equations of hydrodynamics were solved exactly with an EOS of the more general form (1.2'), for two different energies: $\sqrt{s} = 63$ GeV and $\sqrt{s} = 540$ GeV. Three different variants for (1.2') were considered, which all satisfy the following boundary conditions imposed by the assumption of the transition from QGP to hadronic matter: for small energy densities $u \approx 1/\sqrt{7}$ and for large energy densities $u \approx 1/\sqrt{3}$ [h]. The results for the rapidity and low transverse momentum distributions obtained with these variants were compared among themselves, as well as with those derived for a constant speed of sound $u = 1/\sqrt{3}$. It turned out that the four different equations of state lead to quite similar results, all in rather good agreement with data. This was not the case if one considered another constant value for u, e.g. $u = 1\sqrt{4}$, which shows the sensitivity of the physics on the value of u at high energy densities.

---

[g] For simplicity we omit the second independent thermodynamical variable.
[h] As far as I can gather Ref.[13] is the first application of the hydrodynamical model in which a realistic EOS with a variable velocity of sound was used (one of the variants considered included also the influence of confinement on the QGP EOS). This procedure has now become standard in heavy ion physics.

The above results strongly suggest that in the hydrodynamical expansion with a variable velocity of sound only the initial value of u matters and this explains the success of Landau's educated guess [i].

Independent of this interesting historical background, <u>we have to consider the results presented above at their face value: as evidence for QGP</u>.

*Large transverse momenta*

Further support for this conclusion comes from large transverse momenta measured at the CERN intersecting storage ring ISR. While the exponential spectrum describes extremely well the transverse momentum distributions up to $p_T \approx 1$ GeV, this is not the case at larger values of $p_T$, where deviations from this form were observed [j], [14]. Interestingly enough, these deviations, which correspond to the high temperature domain, not only do not contradict the Landau model, but can be used to obtain information about the EOS. Actually they were interpreted as "evidence for a change with temperature of the velocity of sound in hadronic matter and of a phase transition from a strongly interacting hadron phase to a weakly interacting QCD phase" and published under this title in Ref. [15].

The experimental data are represented in Figs. 1 and 2. They show a striking deviation from exponentiality beyond 1 GeV/c; the logarithmic slopes significantly increase with s. However beyond $p_T \approx 5$ GeV/c one observes a resumption of the exponential behavior with an (almost energy-independent) slope of 1.3 $(GeV/c)^{-1}$.

---

[i] This is presumably due to the fact that the entropy of the system is fixed before the (adiabatic) expansion of the system starts.

[j] For a recent review of the circumstances under which large $p_T$ physics in particle physics started and its implications for present heavy ion experiments cf. the paper by Tannenbaum.[14]

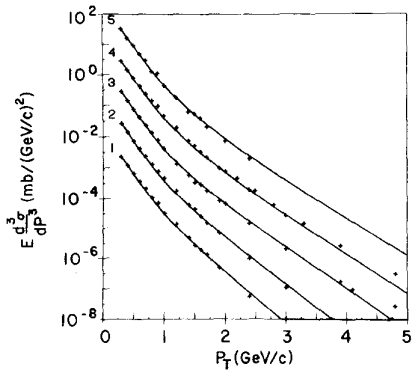

Fig. 1 (From Ref.[15]) Transverse momentum spectra at 90° (c.m. system) for various values of √s (GeV): curve 1, 23 (x $10^{-4}$); curve 2, 31 (x $10^{-3}$); curve 3, 45 (x $10^{-2}$); curve 4, 53 (x $10^{-1}$); curve 5, 63. Statistical errors are smaller than the size of data points. The continuous curves are the results of the hydrodynamical calculations (cf. text below) for $u^2$= 1/6.55.

The interpretation of these experimental results proposed in Ref.[15] starts from the Pomeranchuk [16] freeze-out Ansatz mentioned above, which explains why the bulk of the produced particles have limited transverse momenta ($<p_{T}> \approx 0.3$ GeV/c). However, because of the statistical nature of the process, emission at $T > T_f$ is not absolutely forbidden and this must lead to *leakage* of particles from the excited system, before the expansion has ended and equilibrium has been reached. This preequilibrium emission is known to take place also in medium energy nuclear physics (cf. e.g. [17]). To calculate theoretically this effect Ref.[15] uses the one-dimensional[k] exact solution of the Khalatnikov equation[18] for the relativistic hydrodynamical potential of the Landau model. The functional dependence of the temperature T on time t, viz., T (t) is given implicitly at rapidity y =0 by

$$t(T) = (d/2uw)\{ \int^{\tau} \exp(-w\tau) I_0[(w-1)t'] \, dt' + \exp(-w\tau) I_0[(w-1)] \} \quad (1.4)$$

where d is the proton diameter, $I_0$ is the modified Bessel function, u is the velocity of sound assumed to be constant, $w = (1+u^2)/2u^2$,

---

[k] In the Landau model applied to p-p reactions corrections for the three dimensional motion influence the expansion of the system only at later times. In heavy-ion reactions the situation is different (cf. Section 2.1.2).

$\tau = \ln(T/T_0)$; $T_0$ is the initial temperature given (in units of $m_\pi$) by

$$T_0 = (\varepsilon/\varepsilon_\pi \lambda)^{1/2w} \tag{1.5}$$

with $\varepsilon = E_{cm}^2 w m_\pi^3/\pi m_p$ and $\varepsilon_\pi = m_\pi/V$. $\lambda$ is an integration constant of the equation of state $u^2 = dp/d\varepsilon$.

The invariant cross section $f(p_T)$ reads

$$f(p_T)_{y=0} \sim p_T^{-1} \int dt \int b\, db\, \Phi(p_T, T(t)), \tag{1.6}$$

where

$$\Phi = p_T^2/\{[\exp[(p_T^2 + m^2)^{1/2}/T(t)] - 1\}$$

The integral over t extends from 0 to $t_f$ (the moment of freeze-out defined by $T(t_f) = T_f$) and that over b from 0 to R, where R is the radius of the target. Taking the velocity of sound u and the normalization as free parameters and applying this formalism to CERN-ISR data one finds (cf. Figs. 1 and 2) that:

From $p_T = 0.1$ to ~5 GeV/c the data are well described by the model with a value of u in the range $1/\sqrt{6,4}$-$1/\sqrt{6.8}$, which is compatible with values obtained for u from the hydrodynamical model, when analyzing rapidity distributions in p-p and p-nucleus collisions, and is also in agreement with theoretical predictions.[12] The "large $p_T$" region (1-5 GeV/c) appears as a smooth and natural continuation of the "low-$p_T$" region, and only in the very large $p_T$ regime does something "new" appear:

For $p_T \gg T_o$ (at $\sqrt{s} = 53$ GeV, this happens for $p_T > 5$ GeV/c) equation (3) becomes essentially $f(p_T) \sim \exp(-p_T/T_0)$ and in the region (5-15 GeV/c) the data deviate strongly from this asymptotic form as long as u remains unchanged (~ $1/\sqrt{6.8}$). (See Fig. 2.) However, they are well fitted by an exponential with a higher initial temperature $T_0$ (~ $5m_\pi$, instead of $2m_\pi$) corresponding via Eq. (1.5) to $u = 1/\sqrt{3.5}$. This was interpreted in Ref. [15] as evidence for the fact that the sound velocity u is a (step?) function of temperature and that the high temperature region corresponds to a "different physical situation, probably a new phase". The intermediate $p_T$ region (1-5 GeV/c) could be explained assuming a velocity of sound of $1/\sqrt{6.8}$ with 0.1% of the particles leaked out from a phase with $u = 1/\sqrt{3.5}$ (cf. Fig. 2). Independent support for this conclusion came a few years later (cf. Section 2.2.2 devoted to sQGP) from a calculation of the temperature dependence of the speed of sound, which included the effects of confinement. It was then found that in the temperature range 100-1200 MeV considered the speed of sound varied indeed in the range $1/\sqrt{7}$-$1/\sqrt{3}$ found in Ref. [15]

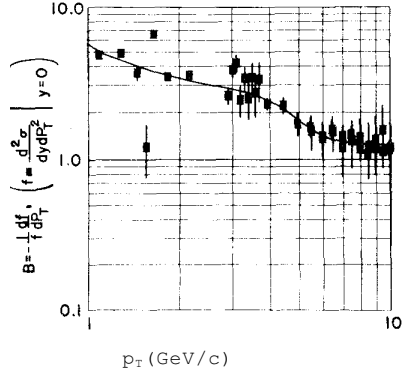

Fig. 2 (from Ref. [15]) Logarithmic slope of the transverse momentum spectra at 63 GeV. The curve is computed from the hydrodynamical model with $u^2 = 1/6.55$ with 0.1% of the particles "leaked" from a phase with $u^2 = 1/3.5$.

As emphasized in Ref.[15] support for the above interpretation comes from the facts that: a) The hydrodynamical model with only one free parameter, viz. u gives a consistent description of the $p_T$ spectra over the whole energy range and in the entire $p_T$ range, for twelve orders of magnitude in cross section. b) This free parameter is already fixed to within a few percent of the fitted value by independent experimental facts like the rapidity distribution, which reflects the freeze out stage and thus corresponds to a value of u in the range $1/\sqrt{6}$- $1/\sqrt{7}$. The energy dependence of the multiplicity, which is determined by the entropy of the initial stage of the system is consistent with a value of $u \approx \sqrt{1/4}$. c)The two values of u can be understood theoretically. The value $u \approx \sqrt{1/6}$ was derived by Zhirov and Shuryak in Ref.[12] for a resonance gas corresponding to hadronic matter and the value $u = 1/\sqrt{3.5}$ follows from lattice QCD calculations, suggesting that the large $p_T$ data from p-p reactions are evidence for a phase transition to quark-gluon plasma.

As far as I can gather Ref. [15] constitutes probably the first phenomenological interpretation of experimental data based on QGP. Its results were used to make predictions for the slope of transverse momentum distributions at higher energies, which could be checked at LHC (cf. Fig. 3) and compared with those derived from a parton model.

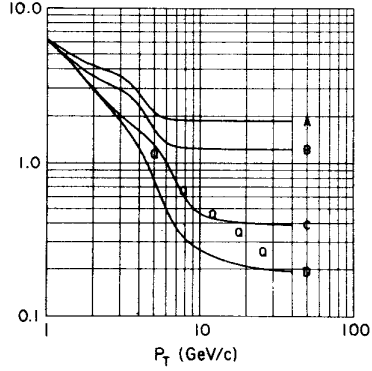

FIG. 3 (from Ref. [15]) . Logarithmic slope of the transverse-momentum spectra: A and B fitted to experimental data at $\sqrt{s}$ =23 and 63 GeV, respectively, C, prediction for $\sqrt{s}$ =1000 GeV assuming the same parameters as at 63 GeV; D, same as C but assuming $u^2 = 1/3$ in the weakly interacting phase. Points labeled Q are from Field's [19] parton-model predictions.

It is interesting that even today the interpretation of large transverse momenta is still an unsettled problem: While the parton model based on hard QCD scattering predicts at large transverse momenta a $p_T^{-4}$ dependence for the invariant distribution, the data show instead rather a larger power ($p_T^{-8}$?) dependence. The truth might lie in the middle: both leakage [l], [20] and parton scattering contribute and the superposition[m] of a power of -4 with an exponential with a small weight might be responsible for an effective power, since leakage is a small effect. As a matter of fact, a detailed consideration (Ref. [21]) of the interplay between the one and three dimensional expansion within the hydrodynamical model shows that in the intermediate $p_T$ region, between 1 and 5 GeV/c, the $p_T$ distribution can be approximated by a power of the order of –5. For a recent discussion of intermediate $p_T$ spectra in terms of a recombination of thermal and shower partons cf. the review by Hwa. [22]

*1.2.3 Multiplicity distributions*

Another type of evidence for quark-gluon plasma in particle reactions based on the equation of state and the Landau model comes from multiplicity *distributions*. The energy dependence of the *mean* multiplicity of produced secondaries has constituted one of the first indications that the equation of state of hadronic matter in its initial stage corresponds to an ideal gas. Once reliable information about the fluctuations of multiplicity, reflected in the multiplicity distributions, became available it was natural to ask whether these distributions supported this conclusion about the equation of state. The answer is positive and can be resumed in the following way.[23] The observed dependence of multiplicity distributions on energy and shifts of rapidity bins can be explained by assuming the existence of two sources: one source is concentrated at small rapidities, has properties of a thermally equilibrated system, as could be expected from a quark-gluon plasma and gives rise to a distribution of negative binomial form $P_1(n_1)$. The other one is contributing to the whole rapidity region, displays characteristics of bremsstrahlung emission and has thus the form of a Poisson distribution $P_2(n_2)$. With increasing energy the weight of the "thermal" source increases. This picture allows a consistent description of the multiplicity distributions in the whole rapidity range as well as in restricted rapidity windows and it accounts for the dependence of the distributions on energy and on the position of the rapidity bins.[24] The justification of these conclusions is given below.

---

[l] In a certain sense the two source core-halo model of Csörgo and Lörstad [20] could be considered as an effective implementation of the leakage phenomenon.

[m] A superposition between a soft and a hard process is also suggested by the RHIC jet quenching data.

*Two types of sources*
According to quantum statistics (cf. e.g. [25]) the forms of multiplicity distributions are bounded by the following two extremes: negative binomials, $P_1(n_1)$, corresponding to chaos and Poisson distributions, $P_2(n_2)$, corresponding to order (coherence). In Ref.[23] one assumes that each multiparticle event is characterized by two types of sources corresponding to these two extremes and that the partition of energy between the two sources is independent of the total energy available for particle production W= K $\sqrt{s}$; K is the inelasticity and
$W = W_1 + W_2$, where $W_1$ and $W_2$ are the energy contents of the two sources.
Experiment shows that on the average, in each multiparticle hadronic event, there exists one leading particle and a central blob. In the generally accepted lore of hadronic collisions in terms of quarks and gluons the leading particles are due to the through-going quarks, while the central blob is formed by the interacting gluons. Such a model can explain among other things the leading-particle effect and reproduces the inelasticity distribution and its energy dependence. [26]

In Ref. [23] it is assumed that the through-going quarks *independently* radiate gluons, part of which hadronize directly. If the hadronization process does not change the form of the gluon multiplicity distribution, (a common assumption in this field) the distribution of these hadrons should therefore be of Poisson-type $P_2(n_2)$. The rest of the gluons as well as the primordial gluons equilibrate. This component is most naturally described by a thermal (negative-binomial) distribution $P_1(n_1)$.

The above conjectures are summarized in the following formulae: total multiplicity $n = n_1 + n_2$ ; total multiplicity distribution $P(n) = \Sigma P_1(n_1)P_2(n_2)$, chaoticity $p = <n_1> / <n>$.

The results of the application of this formalism to proton-proton and antiproton-proton data in the range $\sqrt{s}$ =23-900 GeV are represented in Figs. 4 , 5 and they confirm the properties mentioned above: the existence of two subsystems, due to two sources: The first source is concentrated at small rapidities y and its mean multiplicity $<n_1>$ behaves like $\sqrt{W_1}$. This energy dependence is what one would expect in a Landau-type hydrodynamical approach from a thermally equilibrated source with an equation of state corresponding to an ideal gas, and in particular to a non-interacting quark-gluon plasma. Moreover its rapidity distribution also resembles pretty much what follows from the Landau model. The chaoticity increases with energy from 40% at $\sqrt{s}$ = 30 GeV to 80% at $\sqrt{s}$ = 900 GeV. The second source populates mainly large y and its mean multiplicity $<n_2>$ behaves like $\ln W_2$, a dependence characteristic for a coherent emission mechanism.

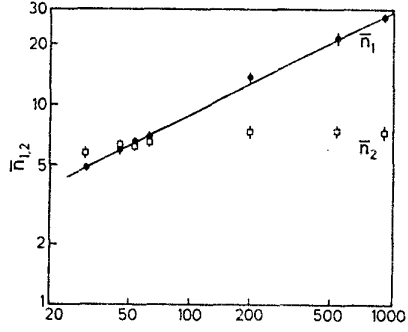

Fig. 4 Dependence of the estimated mean multiplicities $\langle n_1 \rangle$ and $\langle n_2 \rangle$ of the two components on $\sqrt{s}$ (expressed in GeV). The chaotic component $\langle n_1 \rangle$ is represented by circles; the coherent component $\langle n_2 \rangle$ (squares) is shown here on the same log-log scale as $\langle n_1 \rangle$ for the sake of comparison (from Ref. [23]).

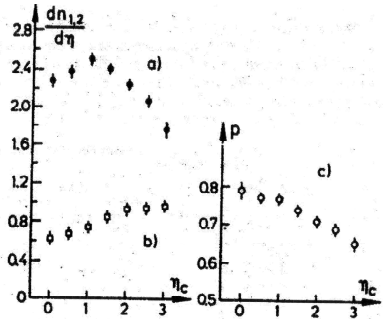

Fig. 5 (a), (b) Dependence of the estimated mean multiplicities $\langle n_1 \rangle$ and $\langle n_2 \rangle$ on the center $\eta_c$ of pseudorapidity windows. (c) Dependence of the chaoticity p on $\eta_c$ (from Ref.[23]).

*Energy dependence of the chaoticity* From Fig. 4 follows that the mean multiplicity of the chaotic source increases (much) faster than that of the coherent one. This means that the chaoticity p increases with energy. This conclusion is supported by an analysis[27] of a different type of multiplicity distributions: While the data used in Ref.[23] referred to the change of multiplicity distributions P(n) with the shift of the center of the (pseudo)rapidity window at fixed width Y of the window, there exists also data about the change of P(n) with Y, the center of the interval being kept fixed. These independent data also convincingly showed that p increases strongly (by a factor of 2) in the energy range $\sqrt{s}$ = 24-540 GeV.

A possible interpretation of this result as suggested in Ref.[23] is the following: In quantum statistics the chaoticity p is an order parameter, which controls the approach of a phase transition; at p = 0 we have a completely coherent phase, while at p = 1 a completely chaotic one. If the system is in thermal equilibrium the first phase could correspond to a condensate and the second phase to a quark-gluon plasma. In that case the increase of p with energy reflects the "melting" of the condensate with the increase of energy/temperature and the approach to QGP. A possible candidate for the condensate is superfluid hadronic matter for which there exists independent evidence [28] (cf. also Section 2.2.2).

*1.3 Global equilibrium: the universality of the hadronization process*

If "prematter" as defined by Carruthers really exists then it should manifest itself not only in reactions induced by hadrons, but in other hadron producing reactions as well. This has indeed been rather convincingly proven in $e^+$-$e^-$ reactions. For this purpose the energy available for hadron production in p-p reactions has been compared with that in $e^+$-$e^-$ reactions, by taking into account the leading particles effect in the former and the jet structure in the latter. Once this is done it appears that one and the same hadron production mechanism is at work. This strongly suggests that although the initial state is different in the two cases, there exists a universal intermediate state[n], common to both systems. The universality manifests itself, among other things, in the energy dependence of the mean multiplicity, in the inclusive distributions and last but not least in the ratios of particle species of secondaries . Not only are these ratios, considered separately for hadron-hadron and $e^+$-$e^-$ reactions, in accordance with what one would expect from systems in thermal and partial chemical equilibrium at a given freeze-out temperature, but this freeze-out temperature is a universal constant, independent of the initial energy and the initial type of reaction. It is not difficult to guess what this universal intermediate state is made of: quark-gluon plasma.

---

[n] This universality was observed much earlier in purely hadronic reactions, when comparing nucleon-nucleon with meson-nucleon reactions. In the case of $e^+$-$e^-$ reactions it answers also the objection of some critics of the application of statistical methods to these reactions, who argue that here the produced particles had no chance to interact and therefore equilibrium could not be reached: The point is that equilibrium had been reached at the QGP level.

*1.3.1 Inclusive distributions*

With the paper [29] "Evidence for the same multiparticle production mechanism in p-p collisions and $e^+$-$e^-$ annihilation" Basile et al. started a series of reports, in which strong similarities and analogies between these two types of reactions were documented. The key which permits the comparison of the two reactions consists in eliminating their specific differences: p-p and nucleon-nucleon reactions, in general, are characterized by the existence of a leading particle, which takes, on the average, half of the energy available for hadron production[o]. This suggests that the reactions p-p and $e^+$-$e^-$ should be compared, not at the same incident energy $E_{cm}$, but at the same energy available for hadron production, which in p-p reactions is $E_{had} = E_{inc} - E_{lead}$, and in $e^+$-$e^-$ reactions $(\sqrt{s})_{e+e-}$. $E_{lead}$ is the energy of the leading particle. Introducing the fractional variable $p/E_{had}$ Basile et al. showed that the p-p and $e^+$-$e^-$ inclusive distributions in this variable were quite similar. In Ref. [30] this similarity was observed also in the dependence of the multiplicity on $E_{had}$. (Of course, there are limitations in this (p-p) – ($e^+$-$e^-$) analogy. Thus e.g. the similarity in multiplicity distributions applies only to the first moment of the distributions, i.e. to the mean. Furthermore, the transverse momentum distribution in p-p reactions is in a first approximation energy independent while in $e^+$-$e^-$ reactions it is not.)

Last but not least, this universality mechanism seems to hold also in A-A collisions, provided one scales the nuclear data by the number of participants. [31] This supports the idea that the production of QGP is not necessarily limited to the heavy ion reaction domain.

**1.3.2** *Particle ratios*

In thermal equilibrium the ratios of particle multiplicities are determined by their mass according to the formula

---

[o] This „elimination" process gets also support from the two-sources model discussed above, where the through going quarks correspond to the leading particles.

$$N \sim V \int d^3p / [(\exp(p^2 + m^2)/T) \pm 1]$$

where the + sign applies for bosons and the – sign for fermions.
As shown by Becattini [32] this simple prediction is in excellent agreement with experiment for all p-p, p⁻p, π-p, K-p and e⁺-e⁻ hadron production reactions, provided i) the leading particle effect in nucleon-nucleon reactions ii) the jet structure in e⁺-e⁻ and iii) quantum number conservation, and in particular strangeness conservation are taken into account. This last requirement is implemented by multiplying the exponent in the denominator by an effective fugacity γ called suppression factor. One finds that at all available energies the hadron ratios are determined according to the above formula by one and the same temperature T ≈ 170 MeV, strongly suggesting a universal, energy independent freeze out mechanism. Not surprisingly, this value is quite close to the critical QCD temperature, the Hagedorn temperature. On the other hand the volume V increases with √s, again in agreement with the Landau model (cf. Section 1.2.1)[p].

*Heavy Ion Reactions*

The amazing success of the Landau hydrodynamical model in particle reactions, as briefly sketched in the previous sections, is matched by various successes in high-energy nucleus-nucleus reactions. While in heavy ion reactions the issue of the number of degrees of freedom is not directly at stake, there exists at least one prediction for A-A collisions, which lends independent support for the idea that the number of degrees of freedom in *particle reactions* is big enough to justify the application of hydrodynamics. This is so because this prediction follows straightforwardly from the hydrodynamical theory of p-p collisions. Since this prediction is based on the QGP EOS it constitutes independent evidence for QGP in particle reactions.

We refer to the calculation of total multiplicity N for central A-A reactions performed by Landau in his first paper [3] and which was found, already in the 1950's, to be in approximate agreement with experimental cosmic rays data. As will be seen below, this agreement seems to be confirmed by the heavy ion accelerator data.

---

[p] The strangeness suppression factor depends weakly on energy. Actually for the applicability of hydrodynamics only thermodynamical (and not chemical) equilibrium is necessary and even this only on a local scale.

*A dependence of multiplicity*

Recognizing that in p-A and A-A reactions nuclei behave as coherent objects ("It would be completely erroneous to treat such a collision as a series of collisions of nuclear protons and neutrons...") Landau [3] starts from the expression derived for N in p-p collisions

$$N(p) = \kappa \, (E_{lab}/2)^{1/4}$$

where $\kappa$ is a constant. He then argues that the energy density as well as the Lorentz contraction in A-A collisions being the same as in p-p collisions the number of produced particles N(A) is just proportional to the volume of the nucleus, i.e. to the mass number A.. This leads to

$$N(A) = \kappa \, A \, (E_{lab}/2A)^{1/4} = \kappa \, A \, (\sqrt{s}/Am)^{1/4}$$

The characteristic linear A dependence of the total multiplicity, at a given energy per nucleon $\sqrt{s}/A$, predicted by this formula, is in approximate agreement with heavy ion data up to and *including* RHIC energies.[31] More than that could be hardly expected, since at present we know that the above derivation does not take into account, among other things, the leading particle effect in p-p reactions and the equivalent partial stopping in A-A reactions. Moreover the derivation of this A-dependence provides an explanation for some, if not all the scaling regularities observed at RHIC (for a quite related point of view cf. Ref.[33]).

At this point a short comment on the relevance of particle physics as a testing ground for QGP may be appropriate. At a first look it seems that particle reactions present two major disadvantages in this respect:

  a) the number of degrees of freedom;
  b) the possibility to see deconfinement over distances exceeding the nucleon size.

These were the most important reasons why heavy ion reactions have become the main tool in the search for QGP. Unfortunately, despite special signatures designed to prove *extended* deconfinement, this goal has not been convincingly achieved either at the CERN-SPS or at RHIC, and it is questionable whether one might ever be able to demonstrate extended deconfinement in the laboratory by the methods envisaged so far. Surprisingly enough, particle physics itself might help surpass this difficulty. Indeed, if QGP has been seen in p-p reactions and if certain variables (like multiplicity e.g.) in A-A reactions scale with respect to p-p reactions then this strongly suggests that these variables reflect QGP behaviour over nuclear sizes.

It is an even bigger surprise and a strange irony of history that the apparent disadvantage a) turns out to be, against all expectations, rather an advantage in the search for QGP: While in heavy ion reactions the number of degrees of freedom is from the beginning not at stake – the factor A obliges – in p-p collisions it has become an important piece of evidence for the existence of QGP. These considerations confirm the importance of particle reactions in the search for quark matter.

*Traces of QGP in low energy heavy ion reactions?*

It is conceivable that manifestations of quark-gluon plasma have been seen at energies much lower than those of present heavy ion facilities Simple estimates [4] of energy densities suggest that constituent and current quark deconfinement could take place even at Bevalac (1.7 A-GeV) and Dubna (4.5 A-GeV) energies. In the following we will address one particular effect of this kind, because it is also related to the ongoing investigations of elliptic flow at RHIC: we refer to the shear viscosity [q, 34] of matter v. In most applications of hydrodynamics to high-energy reactions viscosity was neglected, i.e. the fluid was considered ideal. The success of hydrodynamics in the pre-RHIC era justifies this assumption (cf. however Ref. [35]). For media with phase transitions finite size and surface effects have to be considered. Considering non-central heavy ion collisions in which the participants may be heated up and undergo a transition to the QGP, while the spectators remain in the hadronic phase, Halzen and Liu [36] suggested that the shear viscosity between the two phases is small. This is so because the exchange of particles between these phases, one of which is in a locally coloured phase, is (almost) negligible and limited to the surface separating the two phases (for a related discussion cf. also Ref. [37]). Solving the equations of hydrodynamics in the transverse plane with viscosity (corresponding to the case when no quark matter is formed) and without viscosity, (corresponding to the case when there is a phase transition) the authors [36] found a *separation in rapidity* of the two phases: in the projectile frame the rapidity distribution of the particles produced from that part of the system, which underwent the phase transition, is shifted towards smaller rapidity values as compared with the distribution of those originating from the part unaffected by the transition.

---

[q] For a study of the effect of viscosity on elliptic flow, albeit within a blast wave model, cf. Ref.[34].

As far as I can gather this effect has not been looked for yet: An obvious difficulty arises from the fact that separation in rapidity is a phenomenon, which could be attributed to other causes as well: Rapidity gaps appear in diffraction scattering and jet production, to mention just two more common effects (for a recent review see . e.g. Ref. 38 ). There exists, however, an alternative signal, also based on the reduced value of the shear viscosity between QGP and hadronic matter, and it is conceivable that this signal proposed in Ref. [38] has been seen.

While Ref. [36] assumes a high degree of transparency, probably not yet reached even at RHIC, Ref. [38] applies to much lower energies, where stopping dominates and where a clear distinction between participants and spectators can be made. This means that in each event there are one or two highly excited fireballs, which contain a lot of baryons and which contribute to the central rapidity region, and - for identical colliding partners - two low-temperature fireballs populating the fragmentation regions. Due to fluctuations the phase transition does not take place in each event. In those events where it does not occur, viscosity contributes to the heating of the system and the "low" temperature of the fragmentation region will be higher than in those events where quark matter is formed. It is amusing to mention that such two temperature events have indeed been seen in two independent emulsion experiments [39, 40] with 1.7 A-GeV Fe beams and subsequently also [41] with 4.5 A-GeV $^{12}$C beams. In these experiments the transverse momentum spectrum of α particles in the projectile fragmentation region was measured. In Ref. [39] it was found that this spectrum is characterized by two effective temperature of 10 and 40 MeV respectively. This was confirmed in Ref. [40] with the important specification that these different temperatures belonged to different events, which was interpreted as evidence for different reaction mechanisms.

## 2. Evidence for QGP in the SPS-RHIC Era

### 2.1 *Implications of Observations at SPS for RHIC*

Apart from these early evidences for QGP obtained in the Pre-SPS Era and apart from the A dependence of the multiplicity mentioned above, which still awaits a more systematic investigation at RHIC, but which so far confirms the predictions of the Landau model based on a QGP-like EOS, some of the experimental results obtained in heavy ion reactions at the SPS and at RHIC accelerators were used by theorists to derive similar conclusions. Since most of these "derivations" are reviewed in the literature, we will limit ourselves in the following to certain selected topics, which are related to the previous developments, but which in part have apparently been overlooked, although they seem to be essential e.g. for the claim made in Ref. [1] (which remains otherwise in great part unsubstantiated), or for an understanding of some startling experimental observations made at RHIC.
We start with some surprises found in the nineties at SPS from the comparison of the exact solution of hydrodynamics with an EOS containing a phase transition from QGP to hadronic matter. These surprises came about in part due to the fact that at SPS besides single inclusive, also double inclusive distributions and in particular HBT correlations were available for comparison.

2.1.1 *Role of the equation of state in the solutions of hydrodynamics*

The equations of relativistic hydrodynamics being non-linear, their solution is a non-trivial mathematical problem. Exact analytical solutions of the equations of hydrodynamics for arbitrary velocities of sound u exist only for the [1+1] dimensions case, and that only for constant u. (For p-p reactions it is the Khalatnikov solution of Section 1.2.2. For the more complex case of p-A reactions, cf. Ref.[42], where also older references are given.) That is why, starting with Landau himself, analytic approximations were used or, alternatively, exact, but numerical solutions.

Another complication in the application of hydrodynamics is represented by our limited knowledge of initial conditions. This explains why some authors preferred to *interpret* physical observables in terms of hydrodynamical and thermodynamical concepts like flow and temperature, rather than to *derive* them from the equations of hydrodynamics. This approach is obviously unsatisfactory, since it throws the baby out with the bath: with such an approach one cannot obtain information about the initial state and the different phases of the system (i.e. the equation of state) which we are really after.

With the advent of powerful computers exact numerical solutions in [3+1] dimensions (for central collisions cylindrical symmetry reduces the [3+1] problem to a [2+1] problem) have become more and more used and this effort has paid. Among other things, in the period 1989-1996 the Marburg group obtained and applied to O + Au, S+S and Pb-Pb SPS data [43] an exact numerical solution (Hylander) of the equations of hydrodynamics for [2+1] dimensions, *without assuming boost invariance*[r]. The EOS used was taken from lattice QCD (LQCD) results: it contained a phase transition between a hadron gas and a QGP at 200 MeV. It is important to note that, according to LQCD, this QGP is, at the temperatures considered, not yet an ideal gas, the corresponding velocity of sound around 300 MeV still being around $1/\sqrt{4}$ instead of the asymptotic $1/\sqrt{3}$. . This point will become more relevant when discussing at the end of this paper the strongly interacting QGP.

The simultaneous comparison with data of the single inclusive rapidity and transverse momentum distributions, as well as of the Bose-Einstein correlations showed that this EOS <u>could explain all the data, while the same EOS without such a phase transition could not.</u> Many of the experimental features explained by this hydrodynamical approach were actually predictions, which were later confirmed in experiment. Not only is this is one of the earliest pieces of evidence for QGP in heavy ion reactions at SPS energies recorded in the literature, but it is from a certain point of view superior to the other indications mentioned in Ref. [1], because of its EOS relevance. In fact, with the realization that suppression of charmonium states and jet quenching are not sufficient conditions to prove the existence of QGP (cf. the discussion at the of Section 2.2) the above mentioned method based on hydrodynamics appears to be at present, despite its well known difficulties, one of the most reliable ones.

---

[r] Besides that and contrary to more approximate hydrodynamical approaches, the freeze-out process was treated in a relativistically covariant manner.

The fact that Landau hydrodynamics "ruled the waves" not only in p-p reactions, but also in A-A reactions, was for many a surprise, because of the erroneous, but dominant belief in boost invariance, which turned out to be in obvious contradiction with data. But this was not the only surprise: The non-linearity of hydrodynamics compounded by the non-linearity of the EOS produced another surprise, which contradicted naïve intuition and "predictions" based on this intuition.

2.1.2 *Longitudinal versus transverse expansion*

Besides taking into account partial stopping, the initial conditions assumed in Ref. [43] were such that on top of the longitudinal expansion there was, from the very beginning, also a transverse expansion. (This approach differs from the other [2 +1] dimensional hydrodynamics approach used in Ref.[44] and later applied also at RHIC energies [45]: Ref. 45 also does not assume boost invariance in the expansion of the fluid, but, unlike Ref. 44, it does not assume initial transverse flow [s].

For the further discussion we use the standard notations:

---

[s] Given the high sensitivity of the hydrodynamics to the initial condition, it is possible that this difference is relevant among other things for the interpretation of RHIC HBT data. Another reason why Ref. [45] could not solve the RHIC HBT puzzle is presumably the neglect of resonances (cf. below).

We parametrize the BEC correlation function in terms of the difference of momenta of the two particles parallel and transverse to the direction of the total momentum of the pair, $q_{parallel}$, $q_{transverse}$ respectively. Defining $q_{out}$ as the projection of $q_{transverse}$ on the transverse momentum of the pair $K_{transverse}$ and $q_{side}$ as the component of $q_{transverse}$ perpendicular to $K_{transverse}$, and assuming a Gaussian distribution one has

$$C_2 = 1 + \lambda \exp(-\tfrac{1}{2} q_{parallel}^2 R_{parallel}^2 - \tfrac{1}{2} q_{out}^2 R_{out}^2 - \tfrac{1}{2} q_{side}^2 R_{side}^2)$$

where $\lambda$ is a phenomenological "decoherence" parameter and $R_{parallel}$, $R_{out}$ and $R_{side}$ are the radii of the source associated with the momenta defined above.

The comparison between theory and experiment shows that contrary to what was assumed until then, the transverse expansion plays from the very beginning an important role. This is seen in particular in Bose-Einstein correlations, where in opposition to what one might expect intuitively, the transverse flow does not increase the transverse radius[t]; if anything, it rather contributes to its decrease.[46] This is illustrated in Fig. 6 where the second order correlation function $C_2$ is plotted against the longitudinal and transverse momentum difference, respectively, in the case of [1+1] and [3+1] expansion for central S+S 200 AGeV collisions.

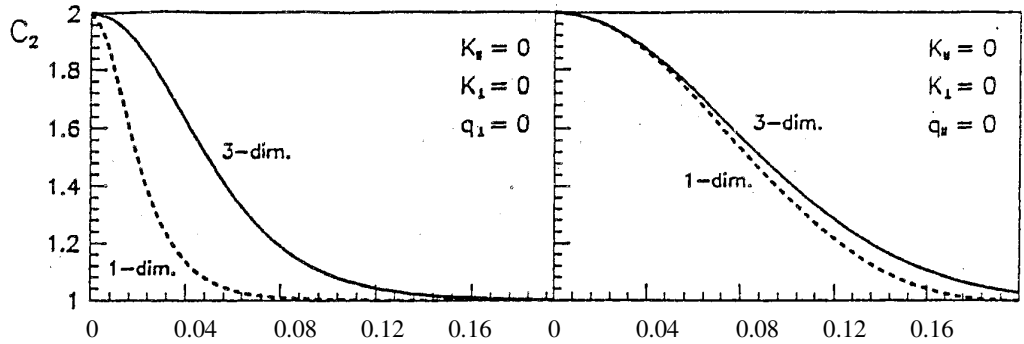

Fig. 6 Second order correlation function in terms of longitudinal (r.h.s) and transverse momentum difference (l.h.s.) (From Ref. [46]).

---

[t] This effect is a consequence of the fact that with increasing time the longitudinal coordinate of the points of freeze-out increases, while their radial coordinate decreases..

*2.1.3 Role of resonances in HBT interferometry; the $R_{out}/R_{side}$ ratio*

Another important conclusion reached in the study of second order correlation functions in heavy ion reactions is that resonances play an essential role. The influence of resonances on Bose-Einstein correlations e.g. can be seen in Fig. 7 , where the contributions of resonances are successively added to the correlation function of direct (thermal) negative pions.

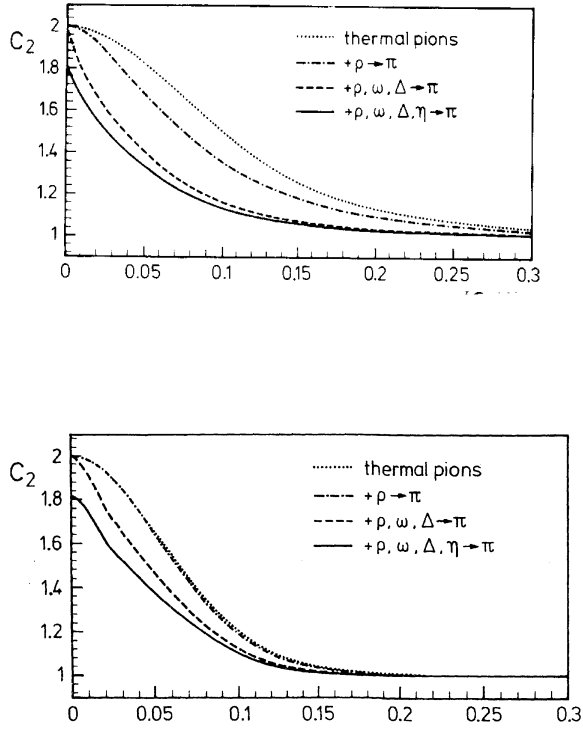

Fig. 7 Bose-Einstein correlation function of negatively charged pions in longitudinal (upper figure) and transverse (lower figure) direction. The contributions from resonances are successively added to the correlation function of direct (thermal) pions (dotted line). The solid line describes the correlation function of all negative pions. (From Ref.[46])

It is observed that the width of the correlation functions and thus the effective sizes of radii progressively decrease as the contributions of longer lived resonances are taken into account. This last fact combined with the interplay between the transverse and longitudinal expansion manifests itself also in the ratio r = $R_{out}/R_{side}$ proposed as a signal for a long-lived QGP:

A hydrodynamical investigation by Rischke and Gyulassy [47], which assumed either boost-invariant or spherical geometry, thus neglecting the interplay between longitudinal and transverse expansion, and which referred only to directly produced pions, lead to the conclusion that for a long-lived QGP the ratio $R_{out}/R_{side}$ increased with $K_{transv}$ and exceeded significantly the value of unity.

On the other hand, the calculations of Ref.[48] for SPS energies illustrated below in Fig. 8 showed that the assumption of a long-lived QGP in itself does not guarantee an increase of r beyond unity: The more general hydrodynamical approach, which does not assume boost-invariance, which considers from the beginning the longitudinal and transverse expansion and which takes into account resonances in the final state, leads to r ≈ 1 *even for a long-lived QGP* [u].

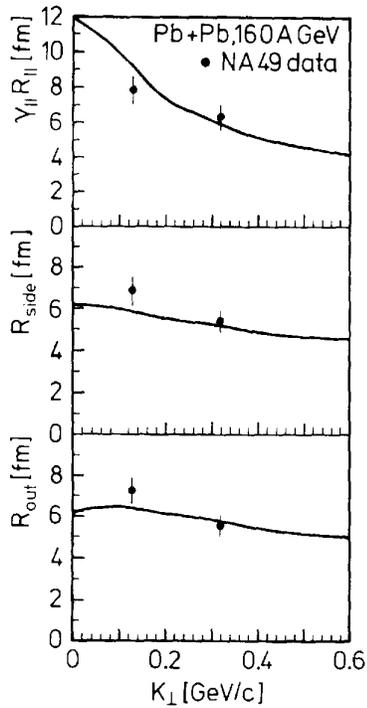

Fig. 8 Effective radii extracted from Bose-Einstein correlation functions as a function of the transverse average momentum of the pair for all pions (from Ref. Ref.[48]).

---

[u] Moreover, this approach permits, among other things, the calculation of the rapidity distributions and of $R_{long}$, quantities, which are beyond the reach of models which assume boost invariance or spherical expansion.

These results clearly showed how some ugly facts like violation of boost invariance and resonances could spoil a nice idea, and this lead to an explicit warning [25,49] against assuming boost invariance and in particular against using [49] r as a signal of QGP.

## 2.2 Surprises from RHIC?

### 2.2.1 HBT puzzle[50]?

This however did not prevent some authors to do exactly that and then to be surprised [v] that the r signal did not deliver what it was supposed to deliver: all RHIC measurements for Au + Au reactions show consistently, both at $\sqrt{s_{NN}}$ = 130 GeV and at 200 GeV, that r ≈ 1.

These HBT results imply that at least one of the assumptions of Ref. [47] – (i) the existence of a long lived QGP and/or (ii) boost-invariant hydrodynamics and neglect of resonances - does not hold at RHIC, either.

In view of the evidence for QGP from various independent experimental and theoretical facts, and given our experience at SPS mentioned above, we think that assumption (ii) is the culprit of all this trouble, the more so that we know now [51] that boost-invariance does not hold even at RHIC.

Sometimes it is argued that for certain calculations it is enough to assume boost-invariance only in the mid-rapidity (-1< y < 1) region. This statement has yet to be proven: In view of the non-linearity of the hydrodynamical equations and of the equation state it seems quite questionable, the more so that the flatness of the rapidity distribution in a restricted mid-rapidity interval follows also from the Landau model, without assuming a decoupling between longitudinal and transverse expansion (cf. Ref. [48]).
*Actually this fact is a concrete example that flatness in rapidity does not imply boost invariance.*

As emphasized by many authors, a correct theory has to explain all experimental facts and must not restrict its applicability to a certain class of phenomena or a limited region of phase space. An attempt to explain

---

[v] This surprise made headlines also in the experimental literature, where it was presented as a proof that current concepts about the space-time evolution of pion sources need to be revised.

transverse expansion without explaining the longitudinal one does not make sense.

Another unjustified prejudice, invoked to explain the HBT puzzle, is the assumption that HBT is only sensitive to freeze-out and not to the initial conditions and therefore by improving just the freeze-out mechanism one might get agreement with data. This assumption is clearly in contradiction with the strong interference between longitudinal and transverse expansion observed at SPS (cf. Section 2.3.1) and there is no convincing reason why at RHIC it should be otherwise.

On the contrary we know that second order correlations are more sensitive to details of the production mechanism than single inclusive distributions [25] and this may be one of the reasons why hydrodynamics with boost invariance can apparently explain elliptic flow phenomena (albeit at the price of assuming very early thermalization times), but fails in the interpretation of HBT data (the "HBT puzzle") Besides that it is conceivable that the errors involved in assuming boost invariance are more important in central reactions than in non-central. It would be quite surprising if hydrodynamics could explain *with the same initial conditions* the data on HBT correlations, which were obtained mostly in central collisions, and the elliptic flow data, which refer by definition to non-central collisions. Along these lines it might be of interest to note that T. Hirano[52], in a [3+1] hydrodynamical study of elliptic flow, finds agreement with data only in the mid-rapidity region. While the author concludes from this that early thermalization can take place only at mid-rapidity, one might consider this result as a further indication that the apparent agreement obtained in explaining certain characteristics of elliptic flow is fortuitous. Actually the early thermalization feature may be a consequence of this difference: In central collisions (almost) all nucleons participate in the equilibration process and this takes more time than in peripheral reactions, where only a fraction of nucleons are participants. Another factor contributing to this difference might be shock waves. Landau hydrodynamics in its original version applies after shock waves have disappeared. In peripheral reactions shock waves presumably play a lesser role than in central collisions and therefore the thermalization process can start earlier.

However the assumption of boost invariance is presumably not the only reason of the HBT puzzle. Another even more probable reason is the neglect of resonances, not only in [50] but also in the other more sophisticated hydrodynamical approaches, like that of Ref. [53], which uses an exact [3+1] solution of hydrodynamics, but where resonances are considered only in single inclusive distributions. Other possible reasons for the difficulties encountered in [53] are the initial conditions which, in opposition to Refs. [46, 48], do not allow for transverse flow from the very beginning, and the EOS for the QGP phase used (cf. below).

To conclude this paragraph: at present it appears unjustified to consider the experimental results at RHIC on $R_{out}/R_{side}$ as puzzling or even surprising. If there have been surprises on this issue then it was the overlooking of the results existing in the literature for more than a decade. (The authors of Ref. [47] themselves were aware of and mentioned the problem of resonances, and in Ref. [54] it was even suggested to use kaons instead of pions, to avoid this problem. That the neglect of resonances might be responsible for the failure to explain the HBT data was conjectured also in Ref. [55].)

2.2.2 *Strongly interacting quark-gluon plasma(sQGP)?*

One of the most startling experimental results obtained at RHIC is without any doubt the *superstrong* quenching of jets.[56] While quenching had been expected and considered by many as a signal of the formation of QGP, it was not expected to be that strong. This fact combined with the robust flow effects observed in the same reactions and which confirmed the hydrodynamical nature of hadronic matter lead some people (cf. e.g. Refs. [57], [58]) to the conclusion that the quark-gluon plasma is, near the phase transition, actually a strongly interacting system, baptized sQGP. This was in contradiction with the expectation, shared by most people, that asymptotic freedom, which implies that at high temperatures quark matter is a weakly interacting plasma, starts immediately after the phase transition. We will come back at the end of the paper to the significance of this result from the QGP perspective. For the moment, though, we will attempt to place it within the historical context. We will show that however "unpleasant" this result might be, it did not come as a complete surprise: It was, in fact, predicted [59] as a consequence of the remnant effects of confinement, which do not vanish abruptly at the phase transition, as a naïve application of the bag model could suggest [w].

To understand how this anticipation came about we have to go back to some phenomenological investigations of hadronic matter, which took place before the idea that at high temperatures/densities matter consists of an asymptotically free system of constituents, was introduced. We shall see that these studies, which refer to superfluidity of hadronic matter, are of current interest also from other points of view.

*Superfluidity and chiral symmetry*

In the early seventies certain phenomenological observations of particle production in strong interactions suggested that hadronic matter had superfluid properties. [60] This conjecture came about by studying, within a statistical approach, peripheral, one pion exchange reactions and trying to use these reactions to get information about the "mesonic cloud" of the nucleon. The spectrum of this cloud turned out to be phonon-like, satisfying thus Landau's criterion of superfluidity.

In this framework it was suggested that Hagedorn's maximum temperature was actually a critical temperature, characterizing a phase transition between the superfluid phase of hadronic matter and another, not yet identified, phase. A few years later, this conjecture was reformulated [28] with the help of the sigma model, in an effective field theoretical approach, incorporating quarks. The Lagrangian of this model reads

$$L = \bar{\psi} \, [i\gamma_\mu \partial_\mu - g(\sigma - i\boldsymbol{\pi}\cdot\boldsymbol{\tau}\,\gamma_5)] \, \psi + \tfrac{1}{2} [(\partial_\mu \sigma)^2 + (\partial_\mu \boldsymbol{\pi})^2] - \tfrac{1}{2} \mu^2 (\sigma^2 + \boldsymbol{\pi}^2) - [\tfrac{1}{2} \lambda (\sigma^2 + \boldsymbol{\pi}^2)]^2$$

where $\psi$, $\pi$ and $\sigma$ are the quark, pion and the scalar sigma fields respectively and g, $\lambda$ and $\mu$ are constants. L is invariant under the SU(2) x SU(2) chiral symmetry group; $\sigma$ and $\pi$ transform as the (1/2, 1/2) representation of the group. The symmetry is spontaneously broken by the non-vanishing vacuum expectation value of the sigma field $<\sigma>_{vac} \neq 0$, which plays the role of the Higgs-Kibble boson. Introducing temperature according to the formalism of field theory at finite temperature one can associate, in the spirit of the Landau theory of phase transitions, the phase where the symmetry is broken with the superfluid phase; it corresponds to temperatures below the critical temperature $T_c$. Above $T_c$ the symmetry is restored.

---

[w] It might be interesting to mention the title of Ref. [59] "How free is the quark-gluon plasma?".

We thus have two phases (a) and (b):

$T \leq T_c$, $<\sigma>_{vac} \neq 0$, $m_\pi = 0$, $m = g<\sigma>_{vac} \neq 0$ (a)

$T > T_c$, $<\sigma>_{vac} = 0$, $m_\pi \neq 0$, $m_\sigma \neq 0$, $m = 0$. (b)

Here $m$, $m_\pi$ and $m_\sigma$ are the masses of the quark, pion and sigma [x] respectively.

*Confinement and asymptotic freedom*

The superfluidity model was formulated *before* the quark-gluon plasma era, but, up to a point, it has strong similarities with the picture, which emerged quite soon, after the conjecture of the existence of a deconfined quark matter phase became a subject of active theoretical and experimental search. Indeed, lattice QCD calculations suggested [61] the existence of three distinct phases: at low temperatures and densities we have hadronic matter, while at very high temperatures/densities we have the deconfinement phase, where there exist only massless quarks and gluons. Between these two phases there exists a third phase in which the quarks are deconfined, but still have mass, and therefore the chiral symmetry is still broken. Moreover, in this intermediate, deconfined phase there still exist pions, which are however massless and therefore have a phonon spectrum. This is presumably the supefluid phase (a) described above. (For further developments of the concept of superfluidity of hadronic matter cf. also Refs. [62].) This suggests that what was termed in the seventies "superfluidity of hadronic matter" also contained hints of a deconfined phase.

To make the phase (b) consistent with what is known now, we have , among other things, to melt away the pions and keep only the deconfined, massless quarks and gluons.

---

[x] This scalar field will be later replaced by the quark-antiquark and gluon condensates.

Moreover, now we know that the truly confined phase is presumably situated at temperatures *below* those corresponding to (a). Last but not least the issue of constituent versus current quarks has yet to be clarified.

Another message of the above scheme was that the masses of the quarks at high temperatures/densities vanish.
This last point lead to the idea [63] that confinement, which corresponds to the opposite - low density- limit, can be described, phenomenologically, by making the mass of the quarks at low densities infinitely heavy. In a certain sense this reflects a kind of Archimedes principle, formulated in a similar context by Pati and Salam: inside the nucleon, due to the high density the quark mass is small, outside it, it is (infinitely) big:

$$m = B/n, \qquad (2.1)$$

where B is a constant, taken in [63] to be the bag constant $B^{1/4} = 145$ MeV, and n the quark number density.
Subsequently this approach was generalized in Refs [59], [64] to include also an effective one-gluon exchange and finite-temperature effects [y].
Eventually an equation of state for the QGP was derived, which included the effects of confinement as defined above and which was used to study the phase transition between that state and nuclear matter, described by an effective finite-temperature mean field EOS. The results of this investigation showed that the effects of confinement persisted *up to much higher* densities/temperatures than those derived without considering this effect: The velocity of sound u, e.g., differs appreciably from its asymptotic value $1/\sqrt{3}$ up to quite high values of temperature; at T = 700 MeV one still found $u = 1/\sqrt{3.5}$. As pointed out in Section 1.2.2 the hydrodynamical interpretation of large transverse momentum distributions [15] lead to quite a similar, if not identical result: the transition from large to small transverse momenta corresponds to a change of speed of sound between $u = 1/\sqrt{3.5}$ and $u = 1/\sqrt{6.8}$. These values, which are still in agreement with present lattice results, show that the QCD-lattice EOS describes the strongly interacting QGP. This EOS was subsequently used in most hydrodynamical

---

[y] For the study of phase transitions this generalization is essential; this explains the difference between the numerical results on critical densities in Ref. [63] and those in Refs. [59, 64].

calculations of the Marburg group; in other words these calculations contained already the sQGP input.[z]

It is interesting to mention that the simplicity of the relation between the mass of the quark and the density, represented by Eq. (2.1), has determined many people to use it in other applications, in particular in the study of strange matter; for an incomplete list cf. Ref. [65]. Given the new findings about sQGP at RHIC one could foresee that this use will be intensified in future phenomenological studies of quark matter. Equation (2.1) could serve as a *theoretical laboratory* for the study of the effect of confinement. The usefulness of such theoretical laboratories was demonstrated lately by the Veneziano formula, initially proposed to describe resonance duality, but which lead unexpectedly to an independent derivation of the Hagedorn exponential mass spectrum and ultimately, and even more surprisingly, to string theory. Eq. (2.1) also describes a dual property, quark duality, relating asymptotically free quarks to confined ones. What new surprises this could lead to is written (perhaps even literally) in the stars.

It has been argued (cf. the Phenix White Paper [2]) that the divergence of the perturbative expansion in QCD also anticipated the sQGP development. In my view this divergence just proofs the limitations of perturbative QCD and nothing more. To derive from this the conclusions which were derived in [59], [64] is a far shot and therefore has not been done, indeed. Neither was this done from general considerations, which predict that in the neighborhood of the critical temperature QGP in not yet an ideal gas – actually the confinement effects predicted by Eq. (2.1) go far beyond the critical temperature – nor from plasmon effects, which modify the gluon mass.

One might wonder whether superfluidity may not provide also an explanation for the apparent ideal character of the hadron-parton fluid seen in strong interactions. (The experimental evidence for this ideal character is based, among other things, on the success of ideal fluid hydrodynamics in its application to multiparticle production phenomena in p-p, p-A and A-A reactions, the possible observation of hot spots [66], which can be produced only if heat conductivity in hadronic matter is small, and last but not least on the strength of elliptic flow seen at RHIC, [57] another of the real surprises of RHIC.) Since a superfluid has by definition zero shear viscosity one might expect that a system, which contains a mixture of a normal fluid and a superfluid is more ideal than the purely normal one. Yet, whether this expectation is also fulfilled for a rotating system as that encountered in elliptic flow, is unclear. In a rotating bucket containing *superfluid helium II* e.g. it is not fulfilled: vortices arise, which make the entire liquid move with the bucket. [67] However another superfluid system made of cold strongly coupled atoms apparently behaves near the Feshbach resonance like a perfect fluid. [68,aa]

---

[z] This is not the case with the calculations by Hirano and collaborators (cf. e.g. Ref. [53]), where the QGP is assumed to be an ideal gas.

To summarize the content of this subsection devoted to sQGP: we arrived at the prediction that QGP may continue to be a strongly interacting system at temperatures much bigger than the critical QCD temperature, starting from the superfluid properties of hadronic-partonic matter, as seen in *particle* reactions. Twenty years later sQGP was experimentally discovered in *heavy-ion physics* at RHIC. This lends independent support to the superfluidity hypothesis and the profound link between high energy particle and nuclear physics.

At this point we might try to answer the question raised in the introduction: what has produced the surprising difference in the assessments of the state of the art in the search for QGP between CERN and Brookhaven? I believe it is first of all [bb], [69] the sQGP finding: Not only was this result surprising, but from a certain point of view it was also embarrassing. Indeed, if the

---

[aa] I am indebted to E. Shuryak for a clarifying correspondence on this point and for communicating some of his results before publication.

[bb] One might invoke also other factors which contributed to this difference of tone: charmonium suppression considered by some people as one of the main proofs of QGP has been recently been contested as a potential QGP signal because charmonium states might survive the QGP phase transition.[57] Furthermore it can be falsified by other effects (for a recent discussion of the way final state interactions can mimic in the "comover model" the effect of deconfinement in J/Psi suppression cf. Ref. [69] .) This last possibility applies also for strangeness suppression, another possible signal of QGP .

matter we see at RHIC is strongly interacting, what is the difference between this matter and conventional nuclear matter? Where is the evidence for a *new* state of matter? Up to now the main qualitative signature of QGP was thought to be its weak interaction and this signature has suddenly been lost. Equally embarrassing might be the repercussions of this state of affairs for the proof of extended deconfinement, another key characteristic of QGP: Given the strongly interacting character of the matter found at RHIC some people are now asking whether the question of extended deconfinement makes even sense for such a system. On the other hand the fact that the strongly interacting character of QGP had been predicted before its experimental observation could serve as an argument in favor of the interpretation of the dense matter found at RHIC as genuine QGP.

*Outlook*

We have seen that in the search for quark gluon plasma, like in any research field, there have been real and less real surprises. For many a surprise may have been that the evidence for QGP is probably much older than usually thought. Another surprise is the breakdown of intuition. To quote just two examples, which illustrate this point: (i) transverse expansion does not increase the transverse radius; (ii) asymptotic freedom, and the fact that the velocity of sound at present energy densities has reached almost the ideal gas limit, do not mean that the QGP is a weakly interacting system. Both these examples are related to hydrodynamics: The first arose from the comparison of the solution of the equations of hydrodynamics with data. The second implies that reliable evidence of QGP can be based only on its definition through the equation of state, which again means comparison of the experimental data with hydrodynamical outputs. The above conclusions can be summarized in the "equation"

$$\text{QGP "=" EOS "=" Hydrodynamics.}$$

This realization is now shared also by experimentalists (cf. e.g. the Phenix White Paper) and, as a consequence of the experimentally established fact that boost invariance does not hold, the Landau hydrodynamical model seems to experience a second resurrection and, most significantly, this time among nuclear experimentalists. [51, 70] One might even assume that hadn't the hydrodynamical calculations (due to unjustified simplifications) encountered difficulties in HBT interferometry, the conclusions of the RHIC White Papers would have been different. Here from follows that exact, albeit numerical, solutions of the equations of hydrodynamics and detailed freeze-out calculations including resonances are at present a must if further progress in this field is to be achieved. The effort and expense implied by this is negligible compared with what has been invested on the experimental side (this refers a fortiori to LHC). Given the precedents and their implications, this task is perhaps too important to be left to theorists only, the more so that at SPS, already, it was found that a detailed knowledge of detectors is necessary for the comparison of theory with data. [cc]

The evidence for QGP in particle physics, which preceded that in heavy ion physics, must not be underestimated: one may question how the entire field of heavy ion reactions would look like and where our present understanding of QGP would be without these early signals of a new state of matter of strong interactions, which have kept alive the interest in this Holy Grail of modern physics. Nevertheless it is difficult to imagine how the important and "novel" realization that QGP is at present energies still a strongly interacting system, could experimentally be seen but in heavy ion reactions. If only for this fact, the effort of heavy ion physics has already paid.

The existence of a dedicated heavy ion accelerator RHIC and in particular of the four dedicated detectors BRAHMS, PHENIX, PHOBOS and STAR, which together cover essentially the entire phase space, a quite unique feature in the history of high energy physics, has already proven its usefulness. Due to it and to the efforts of the physicists involved, the search for quark matter has come of age. This is witnessed, among other things, by the quality of data and by the fact that, wherever there was overlap between the measurements of the four independent groups, the results agreed.

---

[cc] This refers in particular to HBT correlations.

But this is just the beginning. Besides the ongoing heavy ion research, at RHIC a systematic experimental investigation of the properties of QCD matter in *particle reactions* will become possible. So far this type of research had to restrict itself to a limited phase space and often to reduced statistics, because multiparticle dynamics was mostly a by-product of experiments devoted to other specific particle physics topics. In this context it would be very interesting to study experimentally in p-p [71], [dd], [72] and p-A reactions [73], [ee] such matter properties like radial and elliptic flow. The observation of this and similar effects would be useful for a better understanding of the structure of the nucleon, one of the most important unsolved problems of strong interactions. In this way nuclear physics would return part of its debt to particle physics.

## Acknowledgements

An instructive correspondence with F. Becattini, A. Capella, T. Hirano, P. Jacobs, M. Lisa, M. Murray, J.P. Ollitrault, U. Ornik, E. Shuryak, M. Tannenbaum, X. N. Wang and G. Wilk is gratefully acknowledged.
I am also indebted to M. E. Mayer, J.P. Ollitrault, U. Ornik, E. Shuryak, M. Tannenbaum and G. Wilk for a careful, critical reading of the manuscript and many helpful suggestions.

---

[dd] A long time ago Shuryak and Zhirov [71] suggested that matter properties like flow cannot be seen in p-p reactions. Their argument is based on an analysis of $p_T$ distributions of a few types of particles in these reactions, within a simplified hydrodynamical formalism. Whether a more rigorous hydrodynamical treatment, which also takes into account a more complete resonance spectrum than that of Ref.[71] will confirm this conclusion remains to be seen. For a more optimistic point of view on the applicability of hydrodynamics to p-p reactions cf. also Ref. [72].

[ee] It might be useful to remind that certain predictions made in the 1970's for p-A reactions on the basis of the Landau hydrodynamical model and which may provide important information on the equation of state and on the role of shock waves still await to be tested.[42].